\documentclass[aps,prb,preprint,showpacs,superscriptaddress]{revtex4-1}
\usepackage{amsmath}
\usepackage{amsfonts}
\usepackage{stmaryrd}
\usepackage{amssymb}
\usepackage{graphics,epsfig}
\usepackage{subfigure}
\usepackage{bm}
\usepackage{mathpple}
\usepackage{CJK}
\usepackage{color}
\begin{document}
\begin{CJK*}{GBK}{}
\title{Superfluid weight and Berezinskii-Kosterlitz-Thouless transition temperature of strained graphene}
\author{Feng Xu}
\email{xufengxlx@snut.edu.cn}
\affiliation{School of Physics and Telecommunication Engineering,
Shaanxi University of technology, Hanzhong 723001, China}
\affiliation{Institute of Graphene at Shaanxi Key Laboratory of Catalysis, Shaanxi University of technology, Hanzhong 723001, China}
\author{Lei Zhang}
\affiliation{School of Physics and Telecommunication Engineering,
Shaanxi University of technology, Hanzhong 723001, China}
\affiliation{Institute of Graphene at Shaanxi Key Laboratory of Catalysis, Shaanxi University of technology, Hanzhong 723001, China}
\author{Liyun Jiang}
\affiliation{School of Physics and Telecommunication Engineering,
Shaanxi University of technology, Hanzhong 723001, China}
\affiliation{Institute of Graphene at Shaanxi Key Laboratory of Catalysis, Shaanxi University of technology, Hanzhong 723001, China}
\author{Chung-Yu Mou}
\email{mou@phys.nthu.edu.tw}
\affiliation{Physics Division, National Center for Theoretical Sciences, P.O.Box 2-131, Hsinchu, Taiwan, R.O.C.}
\affiliation{Center for Quantum Technology and Department of Physics, National Tsing Hua University, Hsinchu 30043, Taiwan, R.O.C.}
\begin{abstract}
We obtain the superfluid weight and
Berezinskii-Kosterlitz-Thouless (BKT) transition temperature for highly unconventional superconducting states with the coexistence of chiral d-wave superconductivity, charge density waves and pair density waves in the strained graphene.  Our results show that the strain-induced flat bands can promote the superconducting transition temperature approximately $50\%$ compared to that of the original doped graphene, which suggests that the flat-band superconductivity is a potential route to get superconductivity with higher critical temperatures. In particular, we obtain the superfluid weight for the pure superconducting pair-density-wave states from which the deduced superconducting transition temperature is shown to be much lower than the gap-opening temperature of the pair density wave, which is helpful to understand the phenomenon of the pseudogap state in high-$T_c$ cuprate superconductors.  Finally, we show that the BKT transition temperature versus doping for strained graphene exhibits a dome-like shape and it depends linearly on the spin-spin interaction strength.
\end{abstract}
\maketitle
\section{Introduction}

Graphene is one of the most exciting novel materials and has become a goldmine for fascinating physics\cite{ref1,ref2}. A large number of exotic superconducting states of graphene, such as the extend $s$-wave superconducting state, the chiral $d$-wave superconducting state, $p+ip$-wave superconducting state, have been proposed theoretically after considering the interaction and correlation effects between electrons\cite{ref3,ref4,ref5,ref6,ref7,ref8}. The chiral superconductivity breaks both time-reversal and parity symmetries, which  is a natural extension of the $d$-wave state on the square lattice to honeycomb lattice due to the sixfold symmetry. The graphene, when it is doped to the level ($\frac{1}{4}$ hole-doping) with the van Hove singularity, is particularly proposed to be a chiral d-wave superconductor\cite{ref9}.

The recent realization of superconducting state in a twisted bilayer graphene has inspired reconsideration of theories about the superconductivity that occurs in dispersionless bands\cite{ref10,ref11,ref12}. It is well known that in 3D, the superconducting critical temperature $T_c$ in the BCS theory depends exponentially on product of the electronic density of states and the strength of attractive interaction in conventional superconductors. It is then a natural proposal to increase the density of states in the searching of superconductors with higher critical temperatures\cite{ref13}. An extreme route to increase the density of states is proposed by considering pairing in the flat band, which has been shown that in 3D, the superconducting transition temperature in the BCS theory linearly depends on the electron-phonon coupling constant\cite{ref14,ref15,ref16,ref17}. However, since the electronic states in the flat band tend to be more localized, the electron-electron interaction is much stronger in flat-bands. It is therefore necessary to examine how the superconducting transition temperature behaves in flat-bands with strong correlation.

An easy route to realize flat bands in graphene is proposed recently by considering the periodic strain\cite{ref18}.  By considering the strong electron-electron interaction limit, even more exciting, it shows that  the long-sought-after superconducting states with non-vanishing center-of-mass momentum for Cooper pairs (the superconducting pair-density-wave states) can be also realized\cite{ref19,ref20,ref21,ref22}. Nonetheless, since graphene is a two-dimensional matter, instead of being determined by vanishing of the superconducting order parameter, the superconducting transition is determined by the Berezinskii-Kosterlitz-Thouless (BKT) temperature $T_{BKT}$, which is defined as $k_BT_{BKT}=\frac{\pi \hbar^2 n_s}{8m}$, where $n_s$ is the superfluid weight\cite{ref23,ref24,ref25,ref26,ref27,ref28,ref29,ref30}. It is therefore crucial to examine the BKT transition in strained graphene with strong correlation.

In this paper, we investigate the superfluid weight and Berezinskii-Kosterlitz-Thouless (BKT) transition temperature for highly unconventional superconducting states that occur in flat-band of the strained graphene with strong correlation.  We will first compute the superfluid weight and BKT transition temperature for the chiral d-wave superconducting state that coexists with the charge density wave and the pair density wave. We will show that the strain-induced flat bands enhance the  superconducting critical temperature (defined as the BTK transition temperature) approximately $50\%$ in comparison to that of unstrained graphene. Secondly, we compute the superfluid weight of a pure superconducting pair-density-wave state that emerges in the lightly doped graphene and find that the BKT transition temperature is much lower than the temperature when the order of the pair density wave vanishes. This phenomenon is similar to the pseudogap state observed in the cuprate superconductor. Finally, we show that the dependence of the critical temperature on doping exhibits a dome-like shape in resemblance to the superconducting dome observed in high-$T_c$ cuprate superconductors\cite{ref13} and twisted bilayer graphene\cite{ref12}. Furthermore, in the strong correlation limit, the critical temperature depends the residual spin-spin interaction strength linearly. The effect of amplitude and period of strain on the superfluid weight will be discussed at the end. This paper is organized as follows. In Sec.~\ref{sec:theory}, we describe the basic theoretical model and formulation for computing the superfluid weight and the Berezinskii-Kosterlitz-Thouless (BKT) transition temperature for strained graphene. In Sec.~\ref{sec:result},
we present our numerical results and discuss their physical meanings. Finally, conclusions are drawn in Sec.~\ref{sec:con}.

\begin{figure*}
 \includegraphics[width=4.in]{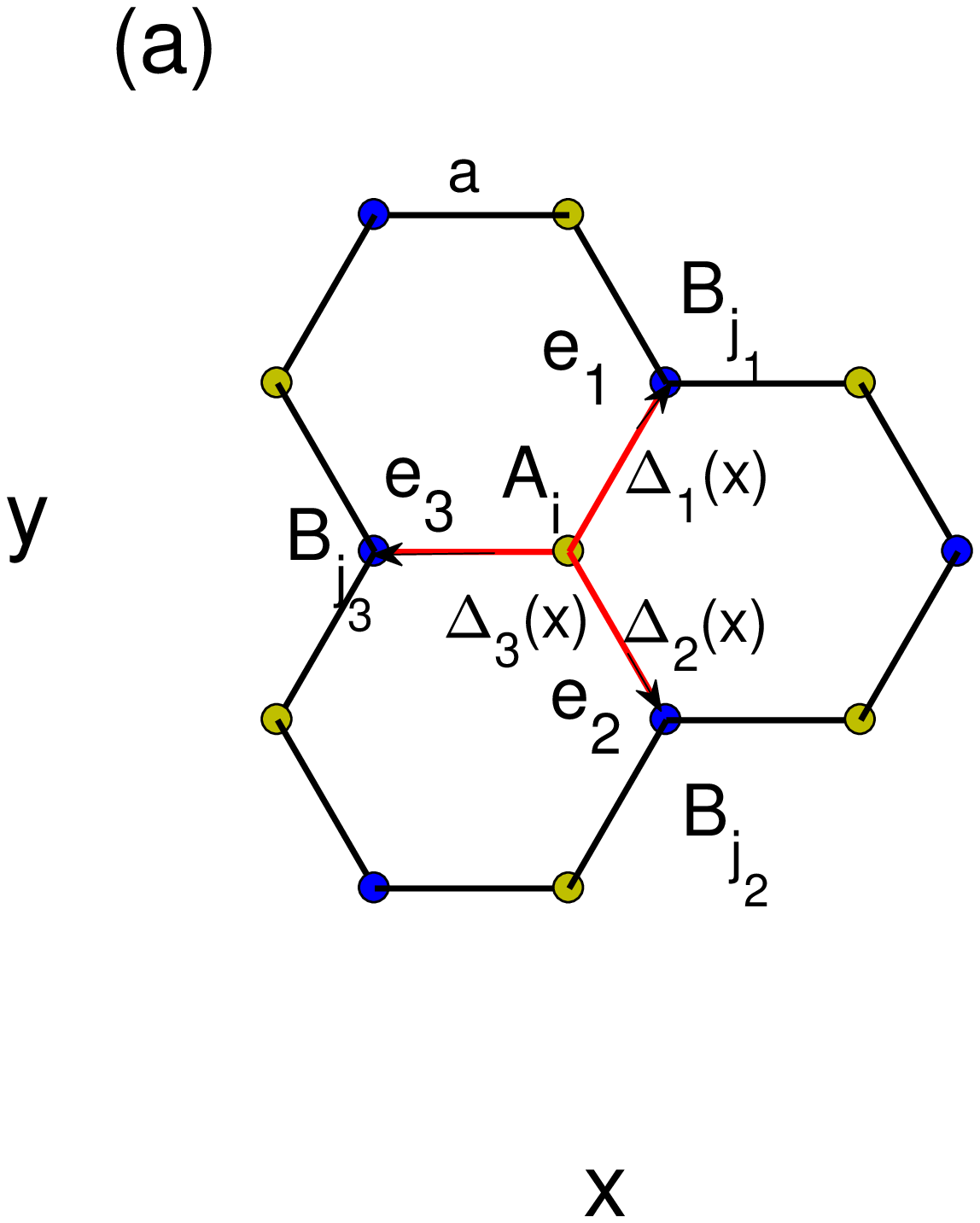}
 \includegraphics[width=2.in]{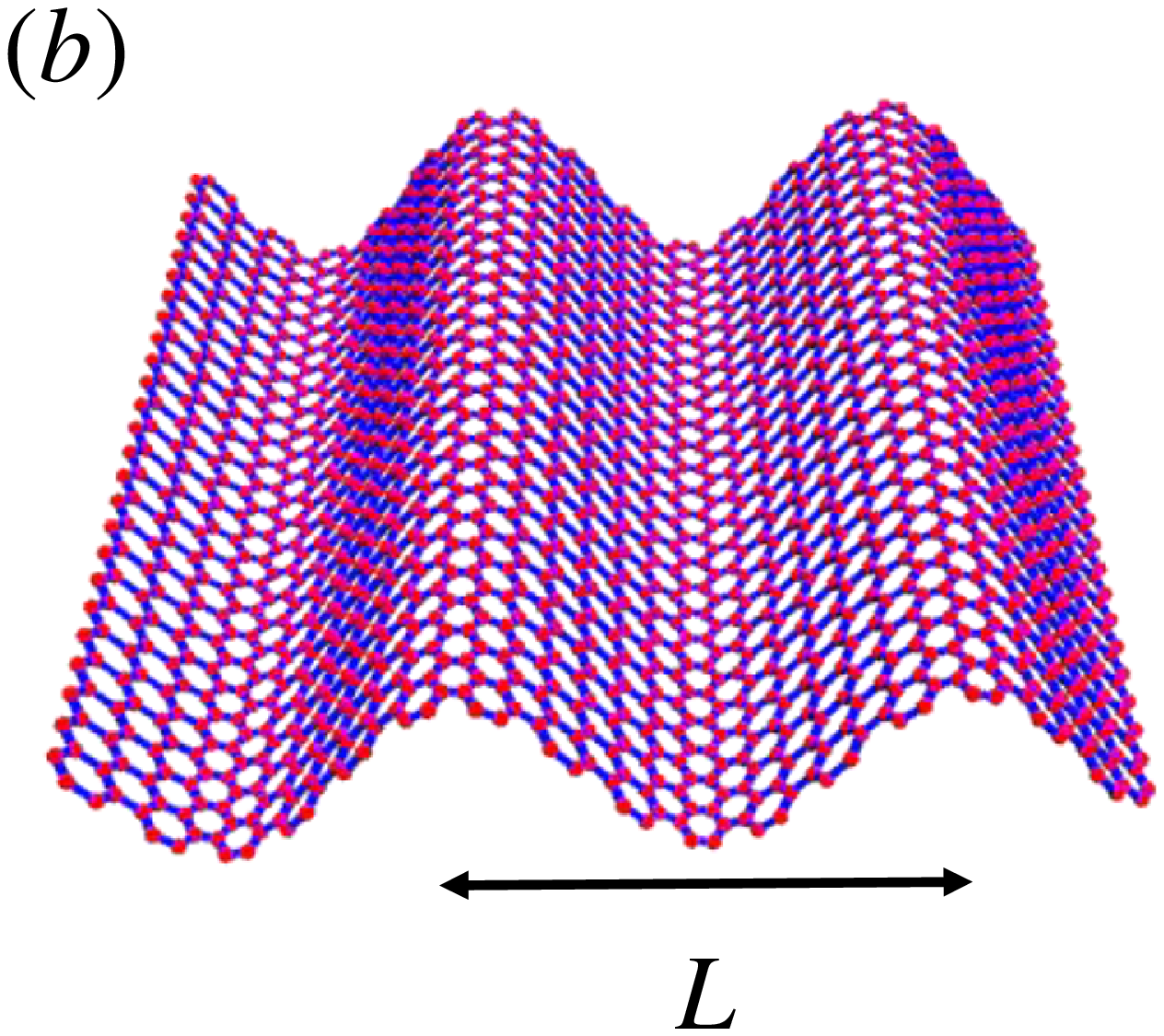}
 \includegraphics[width=3.in]{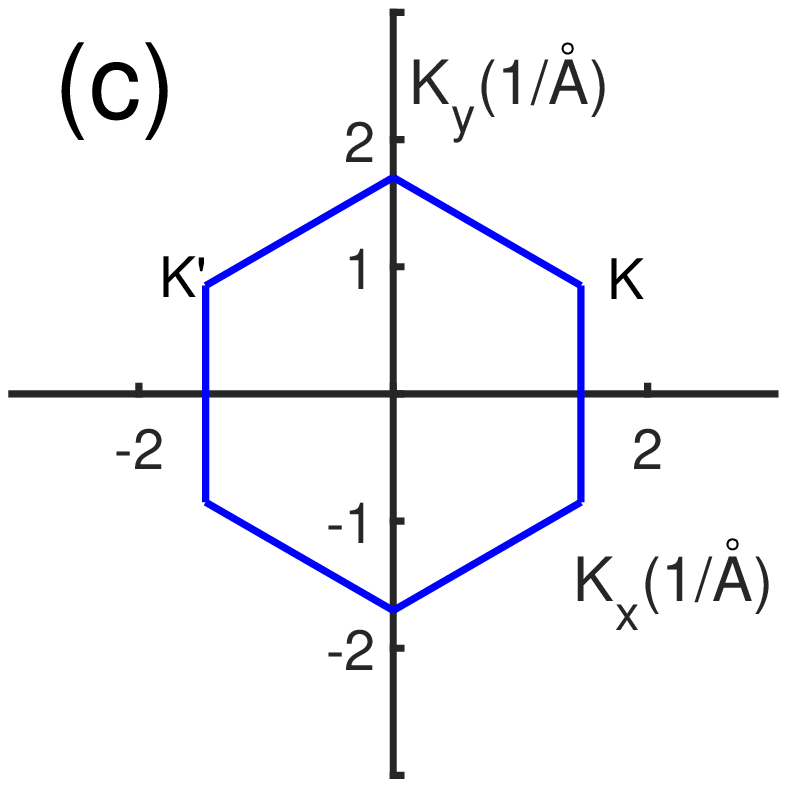}
 \includegraphics[width=3.in]{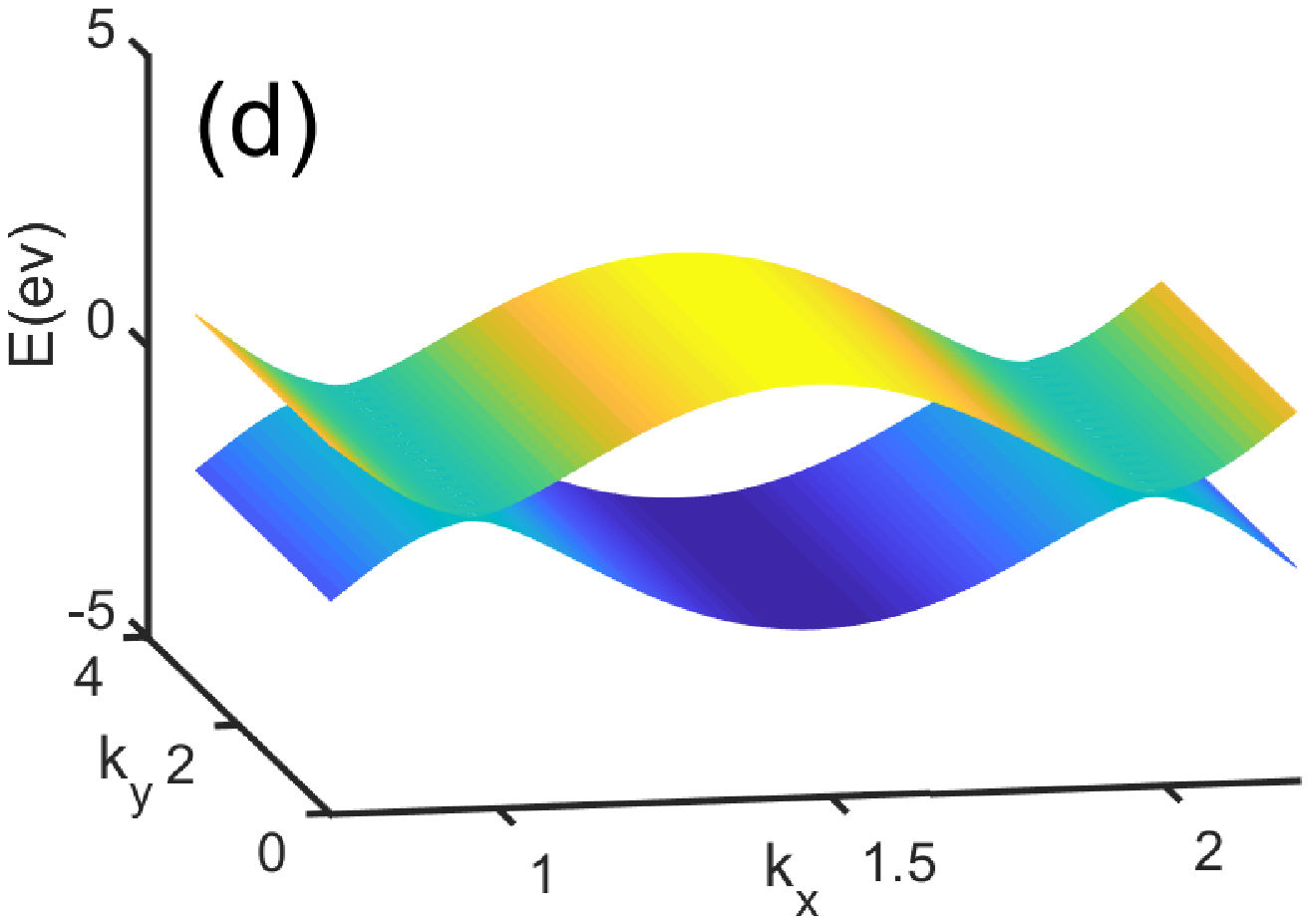}
\caption{Topological flat bands in periodically strained graphene: (a) Honeycomb lattice of graphene, where $e_{i=1,2,3}$ are the nearest-neighbor vectors and $\Delta_{i=1,2,3}(x)$ represents three superconducting orders. (b) Schematic plot of strained graphene. (c) First Brillouin zone of graphene. (d) Lowest energy spectrum of graphene that exhibits flat-bands.}
\label{fig1}
\end{figure*}
\section{Theorerical Formulation}
\label{sec:theory}
The graphene with periodic strain can be achieved in experiments through the presence of ripples with fixed period $L$ or by external stretching as shown in the Fig. \ref{fig1}(b). Theoretically, the strain generally induces changes of hopping amplitudes for electrons. In the tight-binding limit, we shall keep hopping amplitudes to the nearest-neighbors that are characterized by three vectors $\vec{e}_{1}$, $\vec{e}_{2}$, and $\vec{e}_{3}$ as shown in Fig. \ref{fig1}(a),where $\vec{e}_{1}=\frac{a}{2}(1,\sqrt{3})$, $\vec{e}_{2}=\frac{a}{2}(1,-\sqrt{3})$, and $\vec{e}_{3}=-a(1,0)$. The equilibrium hopping amplitude $t=2.8 eV$ is fixed and $a=1.42 \dot{A}$ is the equilibrium bond length. The deformed bond for the strain will be assumed to be the nearest-neighbor bond associated with $e_{3}=-a(1,0)$ and the periodically-modified hopping amplitudes are consistent with the strain period $L=\frac{3na}{2}$,where $n=2,3,4\ldots$. The tight-binging Hamiltonian of periodically strained graphene is thus given by
\begin{equation}
\mathcal{H}_0=-\sum_{i,j_{1},\sigma} t a^{\dagger}_{i,\sigma}b_{j_1,\sigma}-\sum_{i,j_{2},\sigma} t a^{\dagger}_{i,\sigma}b_{j_2,\sigma}-\sum_{i,j_{3},\sigma} (t+\delta t \cos{\frac{2\pi x_i}{L}}) a^{\dagger}_{i,\sigma}b_{j_3,\sigma}+h.c.,
\end{equation}
where $a_{i,\sigma}$($a^{\dagger}_{i,\sigma}$) annihilates (creates) an electron with spin $\sigma$ on site $R_i$ of sublattice A (an equivalent definition for $b_{i,\sigma}$ and $b^{\dagger}_{i,\sigma}$ is used for sublattice B) and $R_{j_k} = R_i+e_k$. The first Brillouin zone of graphene without strain is plotted in the Fig.\ref{fig1}(c). In the presence of strain, the lowest energy spectrum of graphene in the normal state is shown in the Fig. \ref{fig1}(d). Clearly, it exhibits flat bands in the $k_y$ direction. The strong correlation is included by considering the Hubbard interaction between electrons, $\mathbf{U}\sum_{i}n_{i\uparrow}n_{i\downarrow}$. In the strong interacting limit, the singly occupied state can be described by an effective t-J model as follows\cite{ref31,ref32}
\begin{equation}
\mathcal{H}=P_G\left[ \mathcal{H}_0 +\sum_{i,j}J_{ij}\left(\vec{S}_{i} \cdot \vec{S}_{j}-\frac{1}{4}n_{i}n_{j}\right)\right]P_G,
\end{equation}
where $J_{ij}=\frac{4t^2_{ij}}{\mathbf{U}}$ is the antiferromagnetic (AF) coupling, $\vec{S}_i$ and $n_i$ are the spin and number
operators for electrons at $i$ site respectively, and $P_G = \prod_{i}(1- n_{i\uparrow}n_{i\downarrow})$ is
the Gutzwiller projection operator that projects out states with doubly-occupied sites.  Note that $J_{ij}$
 acquires the spatial dependence through the deformed hopping amplitude. The parameters we shall use in this paper are as follows: $\mathbf{U}=4t$, $J=\frac{4t^2}{\mathbf{U}}$, $t=$ 2.8 eV, the period of strain $n=15$, and the strain strength $\delta t /t = 0\rightarrow0.25$.

To investigate superconductivity of this model, the AF interaction term, $\vec{S}_{i}\vec{S}_{j}-\frac{1}{4}n_{i}n_{j}$, is decoupled into\cite{ref33}, $-\frac{3}{8}(\hat{\chi}^{\dagger}_{ij}\hat{\chi}_{ij}+\hat{\Delta}^{\dagger}_{ij}\hat{\Delta}_{ij})$, where $\hat{\chi}^{\dagger}_{ij}=\hat{c}^{\dagger}_{i\uparrow}\hat{c}_{j\uparrow}+\hat{c}^{\dagger}_{i\downarrow}\hat{c}_{j\downarrow}$ and $\hat{\Delta}^{\dagger}_{ij}=\hat{c}^{\dagger}_{i\uparrow}\hat{c}^{\dagger}_{j\downarrow}-\hat{c}^{\dagger}_{i\downarrow}\hat{c}^{\dagger}_{j\uparrow}$.
We resort to the slave-boson method to investigate this strained graphene model in the mean-field level\cite{ref33,ref34,ref35}. Since the slaved boson condenses in the superconducting state, the electron operator can be written as $c_{i\sigma}=\sqrt{\delta_i}f_{i\sigma}$, where $\delta_i$ is the hole density on site $i$ and $f_{i\sigma}$ is the fermionic spinon operator. The mean-field Hamiltonian is then given by
\begin{eqnarray}
\mathcal{H}_{MF}=\sum_{<ij>,\sigma}-(\sqrt{\delta_i\delta_j}t_{ij}+\frac{3}{8}J_{ij}\chi_{ij})\hat{f}^{\dagger}_{i\sigma}\hat{f}_{j\sigma}-\sum_{<ij>}\frac{3}{8}J_{ij}\Delta_{ij}(\hat{f}^{\dagger}_{i\uparrow}\hat{f}^{\dagger}_{j\downarrow}-\hat{f}^{\dagger}_{i\downarrow}\hat{f}^{\dagger}_{j\uparrow})+h.c.
\end{eqnarray}
Here $\chi_{ij}=\langle \hat{\chi}_{ij}\rangle$, $\Delta_{ij}=\langle \hat{\Delta}_{ij}\rangle$, and the doping level $\delta$ is the average of $\delta_i$. To match with the periodic strain, it is presumed that $\chi_{ij}$ and $\Delta_{ij}$ depend on position as shown in the Fig. \ref{fig1}(b). The order parameters set as
\begin{eqnarray}
&&\Delta_{ij}=\Delta_0+\langle \hat{\Delta}_{\frac{Q}{2}}\rangle \exp{(i\frac{Q}{2}x)}+\langle \hat{\Delta}_{-\frac{Q}{2}}\rangle \exp{(-i\frac{Q}{2}x)}+\langle \hat{\Delta}_{Q}\rangle \exp{(iQx)}+\langle \hat{\Delta}_{-Q}\rangle \exp{(-iQx)},\nonumber\\&&
\chi_{ij}=\chi_0+\langle \hat{\chi}_{Q}\rangle \exp{(iQx)}+\langle \hat{\chi}_{-Q}\rangle \exp{(-iQx)},
\end{eqnarray}
where $Q=\frac{2\pi}{L}$ corresponds to the period of strained graphene in the momentum space. In particular, $\Delta_{ij}$ has a period with $2L$ and when its spatial average vanishes, it corresponds to the form of the superconducting order parameter in the pure pair-density-wave state\cite{ref18}. In the pure pair-density-wave state, we have $\Delta_0=\langle \hat{\Delta}_{Q}\rangle=\langle \hat{\Delta}_{-Q}\rangle=0$ and $\langle \hat{\Delta}_{\frac{Q}{2}}\rangle=\langle \hat{\Delta}_{-\frac{Q}{2}}\rangle\neq 0$, which is similar to  the Larkin-Ovchinnikov state with two complex order-parameters $\langle \hat{\Delta}_{\pm\frac{Q}{2}}\rangle$ and is different from the Fulde-Ferrell state in which there is only one complex  order-parameter field $\langle \hat{\Delta}_{\frac{Q}{2}}\rangle$\cite{ref1r,ref2r,ref3r}.

We perform a discrete Fourier transformation to the
mean-field Hamiltonian with
\begin{eqnarray}
\hat{C}_{i\sigma}=\frac{1}{N}\sum_{\mathbf{k}}\hat{C}_{\mathbf{k}\sigma}\exp{i\mathbf{k}\mathbf{r}},
\end{eqnarray}
the mean-field Hamiltonian can be  then expressed as
\begin{eqnarray}
H_{MF}=\sum_{\mathbf{k},\sigma}\psi^{\dagger}(\mathbf{k})h_{MF}(\mathbf{k}) \psi(\mathbf{k}),
\end{eqnarray}
where $\psi(\mathbf{k})=(C^{\dagger}_{A_\uparrow,k_x+\frac{\pi j}{L},k_y},C^{\dagger}_{B_\uparrow,k_x+\frac{\pi j}{L},k_y},C_{A_\downarrow,-k_x+\frac{\pi j}{L},-k_y},C_{B_\downarrow,-k_x+\frac{\pi j}{L},-k_y}),j=0,1,....,2n-1$. The Hamiltonian are then self-consistently solved with the chiral d-wave superconducting order $\Delta,\exp{(i\frac{2\pi}{3})}\Delta,\exp{(i\frac{4\pi}{3})}\Delta$, hopping amplitude $\chi$, charge density waves in hopping amplitude $\langle \hat{\chi}_{Q}\rangle$, pair density waves $\langle \hat{\Delta}_{\frac{Q}{2}}\rangle$, and $\langle \hat{\Delta}_{Q}\rangle$. The temperature-dependence of a mean-field order $\hat{O}$ is obtained by the follow equation:
\begin{eqnarray}
O(T)=\frac{1}{N}\sum_{k}\langle \hat{O} \rangle=\frac{1}{N}\sum_{nk}\langle \psi_{nk}|\hat{O}|\psi_{nk}\rangle \frac{1}{1+\exp{(\frac{E_{nk}}{k_BT}})},
\end{eqnarray}
where $\psi_{nk}$ and $E_{nk}$ are the eigenfunction and eigenvalue of the mean-field Hamiltonian and $k_B$ is the Boltzmann constant and $T$ is temperature. A flat energy band allows deformations of the particle momentum distribution without energy cost. The pair density wave state has been proposed as an alternative state that gives a lower energy than that  of the conventional BCS
state\cite{ref3r}. The phases diagram about those coexisting states in the ground state and its thermal effect has been discussed  with the Ginzburg-Landau theory in the previous works\cite{ref18,ref22}. Here we focus on the superfluid weight and the Berezinskii-Kosterlitz-Thouless (BKT) transition temperature on this system and the optimal parameters to enhance the superconducting transition temperature.

To determine the superfluid weight, we resort to the linear response theory and compute the current-current correlation function\cite{ref36,ref37,ref38}. The paramagnetic current density ($x$ component) at the position $r$ is defined as\cite{ref36,ref37}
\begin{eqnarray}
J^P_x(r)=it\sum_{\sigma}(\hat{C}^{\dagger}_{r+x\sigma}\hat{C}_{r\sigma}-\hat{C}^{\dagger}_{r\sigma}\hat{C}_{r+x\sigma}),
\end{eqnarray}
and the kinetic energy density associated with $x$-oriented link at position $r$ is given by
\begin{eqnarray}
\hat{K}(r)=-t\sum_{\sigma} (\hat{C}^{\dagger}_{r+x \sigma} \hat{C}_{r \sigma}+\hat{C}^{\dagger}_{r\sigma}\hat{C}_{r+x\sigma}).
\end{eqnarray}
 The kinetic energy density is numerically computed as follow:
\begin{eqnarray}
\label{eq1}
\langle-K_x(T)\rangle=\delta\frac{t}{N}\sum_{nk\sigma} \langle \psi_{nk}|(C^{\dagger}_{B_{k\sigma}}C_{A_{k\sigma}}+C^{\dagger}_{A_{k\sigma}}C_{B_{k\sigma}})|\psi_{nk}\rangle\frac{1}{1+\exp{(\frac{E_{nk}}{k_BT}})},
\end{eqnarray}
where $\delta$ is the averaged hole doping level.
The linear current response produced by the vector potential $A_x(r,t)=A_x(q)\exp{(iqr-i\omega t)}$ is given by
\begin{eqnarray}
\langle J_x(q,\omega)\rangle=-e^2 [ \langle -K_x \rangle-\Lambda_{xx}(q,\omega)] A_x(q,\omega),
\end{eqnarray}
where $J_x(q,\omega)$ is the total current and $\Lambda_{xx}$ is the current-current correlation function for the paramagnetic current density\cite{ref36,ref37,ref38},  which is given by
\begin{eqnarray}
&&J^P_x(k,q)=it\sum_r e^{-iqr}(C^{\dagger}_{j_3}C_{i}-C^{\dagger}_{i}C_{j_3})\nonumber\\&&
=it\sum_k\{C^{\dagger}_{B_{k\sigma}}C_{A_{k+q\sigma}}-C^{\dagger}_{A_{k\sigma}}C_{B_{k+q\sigma}}\}.
\end{eqnarray}
The current-current correlation function is then given by
\begin{eqnarray}
\label{eq2}
&&\Lambda_{xx}(q, i \omega_m )=\delta\frac{1}{N}\int_{0}^{\beta}e^{i\omega_m\tau}\langle J^P_x(q,\tau)J^P_x(-q,0) \rangle\nonumber\\&&
=\delta\frac{1}{N}\sum_k J_x(k,q)J_x(-k,-q)TrG_0(k+q,i\omega_m+i\Omega)G_0(k,i\Omega)(-1)\nonumber\\&&
=\delta\frac{t^2}{N}\sum_{k,n,m}  \mathbf{U}^{\dagger}(k)\Xi \mathbf{U}(k+q)\mathbf{U}^{\dagger}(k+q)\Xi\mathbf{U}(k) \frac{f(E_n(k+q))-f(E_m(k))}{E_n(k+q)-E_m(k)+i\omega_m},
\end{eqnarray}
where $\mathbf{U}(k)=(\psi_{1k},\psi_{2k},...,\psi_{nk})$, $\Xi$ is the matrix form of $J_x$, $f(E_n)$ is the Fermi-Dirac distribution function, and $\omega_m=2\pi m k_BT$ is the Matsubara frequency which will be set to $i \omega_m  \rightarrow  \omega+i \delta$ in the usual analytic continuation. Note that the current-current correlation function is related to the electric conductivity through the relation
\begin{eqnarray}
\sigma_{xx}(\omega)=-e^2\frac{\langle-K_x\rangle-\Lambda_{xx}(q=0,\omega)}{i(\omega+i\delta)}.
\end{eqnarray}
The superfluid weight $D_s$ is then obtained by setting
\begin{eqnarray}
\frac{D_s (T)}{\pi e^2}=\langle-K_x\rangle-\Lambda_{xx}(q_x=0,q_y\to 0,i\omega_m= 0) \equiv  \frac{n_s (T) }{m} \label{ns}
\end{eqnarray}
with the restricted condition $\langle-K(x)\rangle-\Lambda_{xx}(q_x\to 0,q_y= 0,i\omega_m= 0)$ as required by gauge invariance. This superfluid weight definition is equivalent to the one defined via the change of free energy due to the phase twist applied to the superconducting order parameter\cite{ref16,ref6r}.  Here the superfluid density $n_s (T)$ is determined through the superfluid weight as $n_s (T) = m \frac{D_s}{\pi e^2}$ with $m$ being the effective mass for the Cooper pair. By ignoring the anisotropic effect on the superfluid weight and assuming $D_s(x)=D_s(y)$ for simple computation, in fact, the superfluid weight on the $y$ direction can be obtained in the same way.

As the strained graphene is a two-dimensional system, the critical temperature transition to superconductivity is determined by the Berezinskii-Kosterlitz-Thouless (BKT) transition temperature as\cite{ref23,ref24,ref25,ref5r}
\begin{eqnarray}
k_BT_{BKT}=\frac{\pi\hbar^2}{8} \frac{n_s(T_{BKT})}{m}.
\end{eqnarray}
Since the superfluid density $n_s(T)$ is obtained by numerical calculating Eq.\ref{eq1} and Eq.\ref{eq2}, it is equivalent to the definition in terms of the stiffness of the superconducting order parameter in the thermodynamic potentials. The definition of the Berezinskii-Kosterlitz-Thouless (BKT) transition temperature has long applied in the multiband system with great success, such as twisted bilayer graphene with valence electrons and flat bands\cite{ref30,ref6r}.

\section{Results and discussions}
\label{sec:result}
\begin{figure*}
 \includegraphics[width=3.2in]{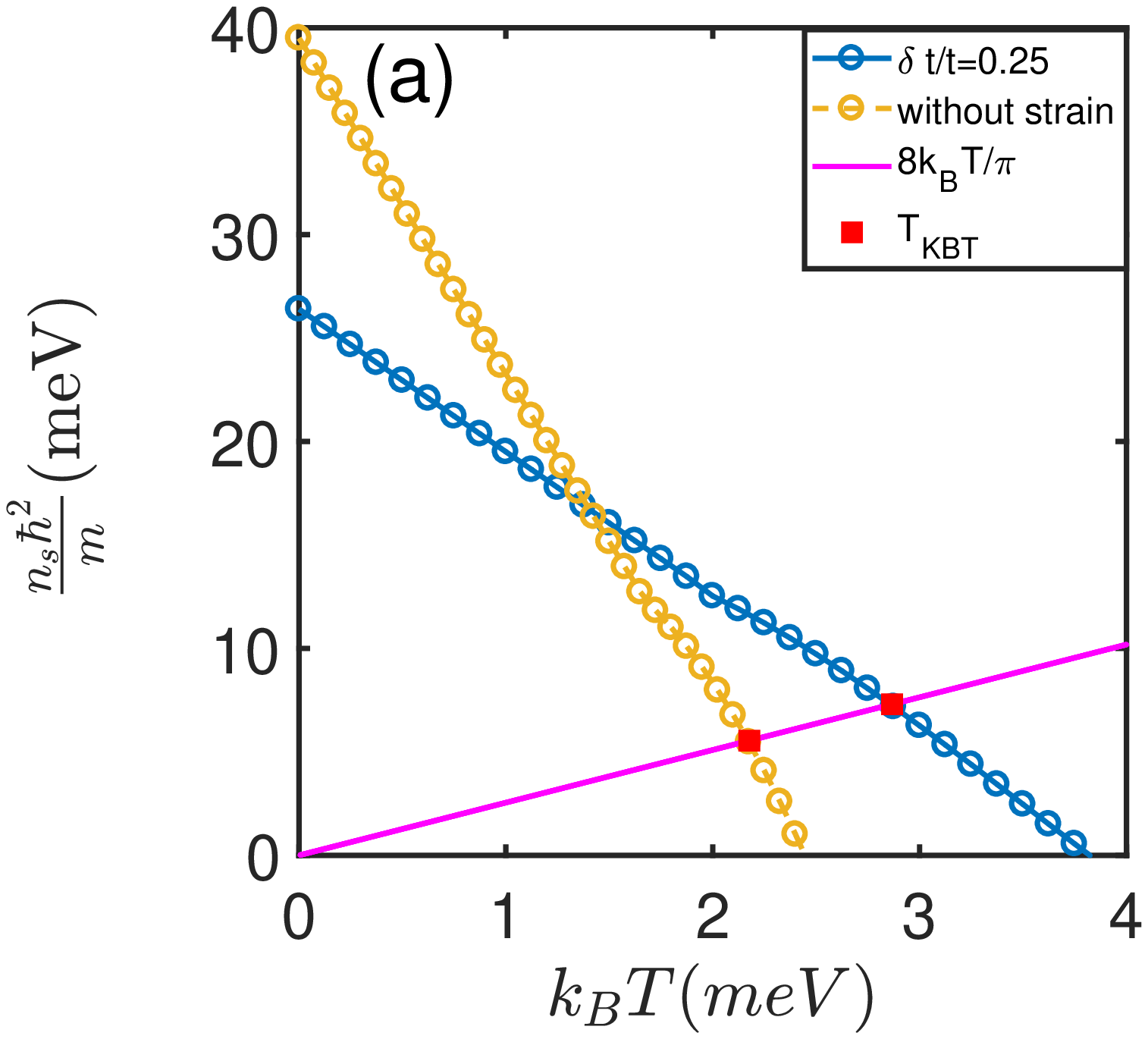}
 \includegraphics[width=3.2in]{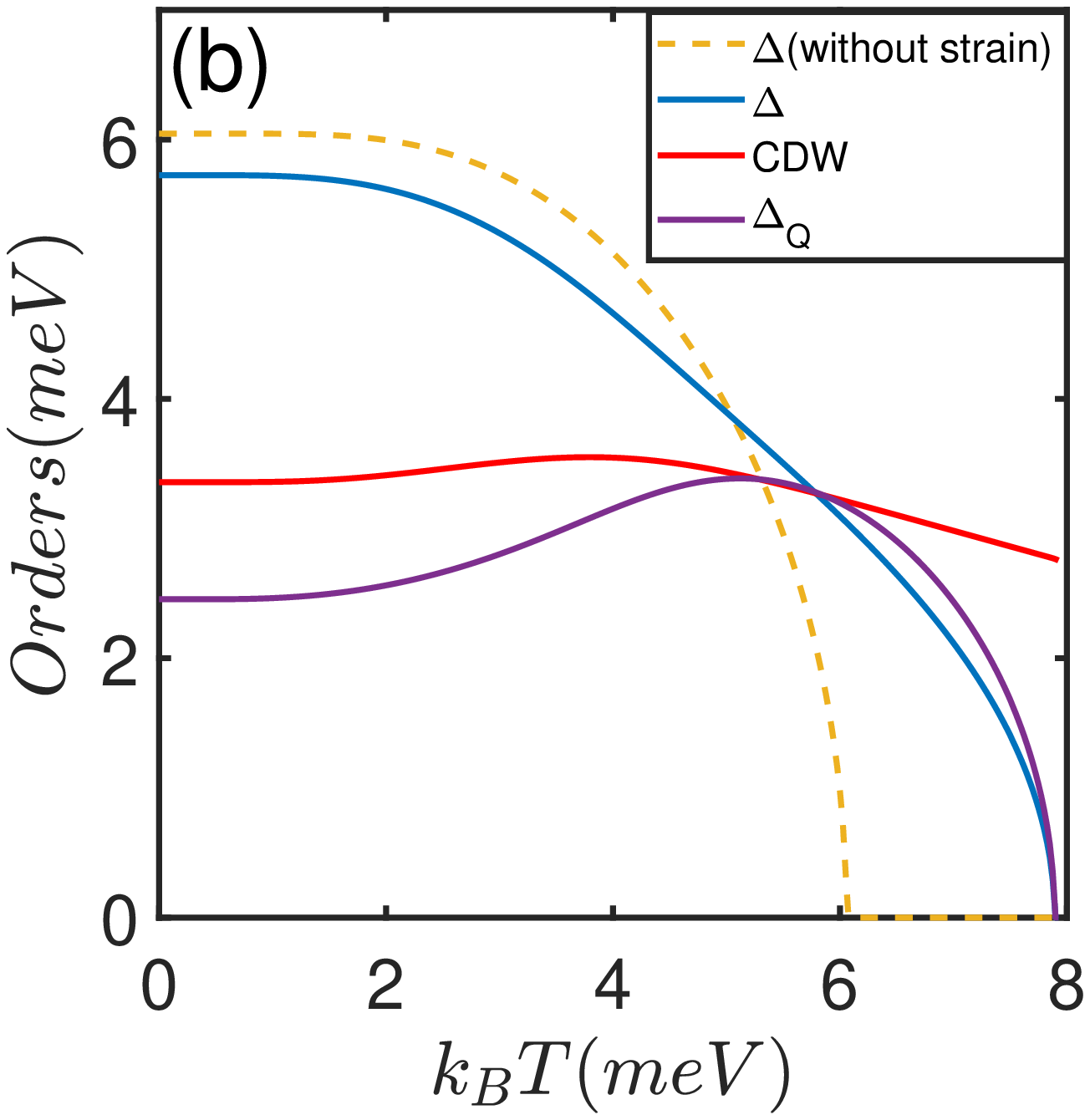}
 \includegraphics[width=3.2in]{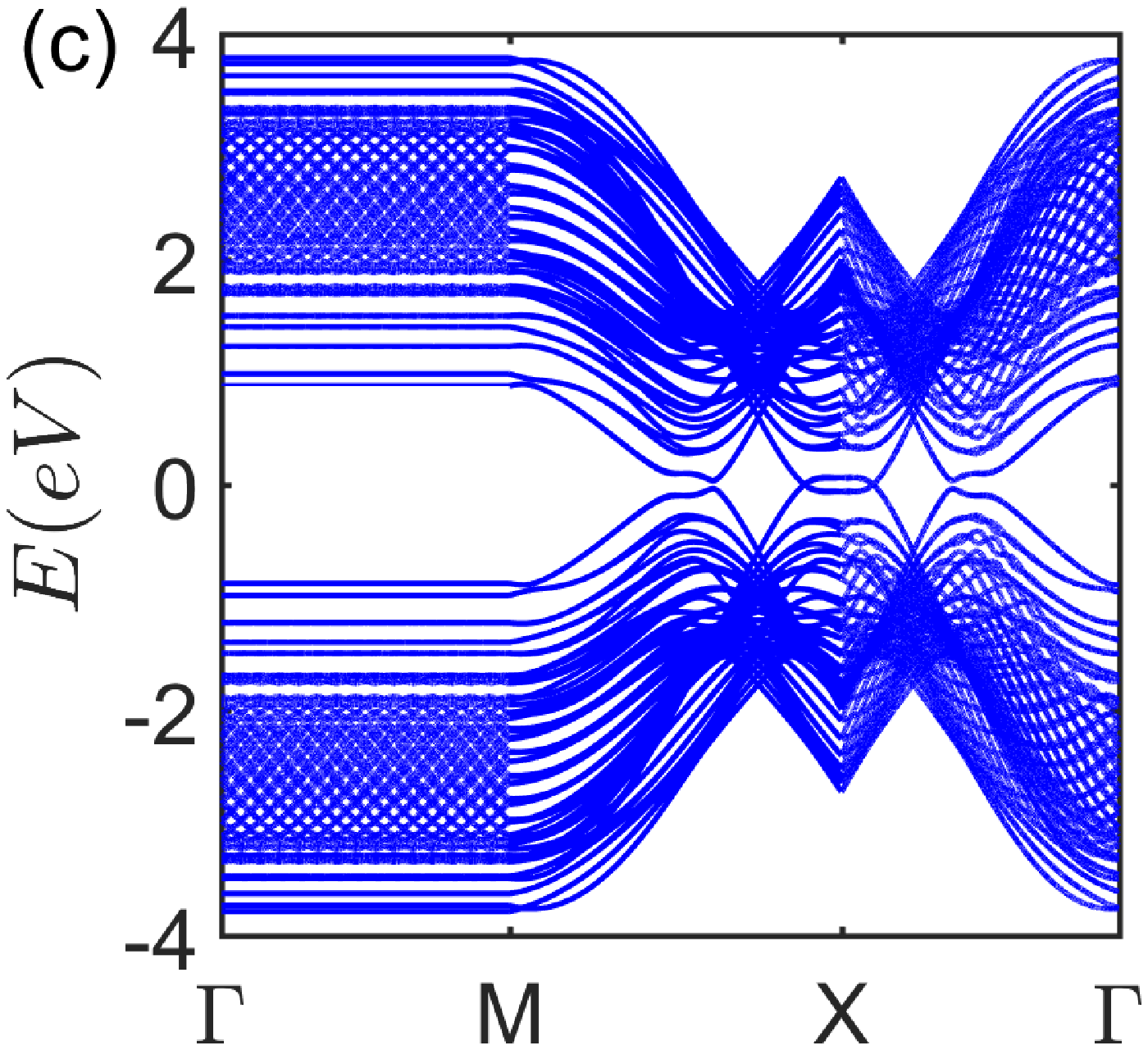}
\caption{(a) The superfluid weight for the $\frac{1}{4}$-hole doped graphene with $\mathbf{U}=4t$ as a function of temperature without or with strain. The Berezinskii-Kosterlitz-Thouless transition temperatures are intersections of $\frac{\hbar^2 n_s(T)}{m}$ with $\frac{8 k_BT}{\pi}$ as indicated by red dots. (b) The chiral d-wave superconductivity order, pair density wave order with momentum $Q$, and charge density wave order versus temperature for the $\frac{1}{4}$-hole doped graphene with strain amplitude being $\frac{\delta t}{t}=0.25$ and period being $n=15$. The pure chiral d-wave superconductivity order in the doped graphene without strain is shown as the dash line for a comparison.  (c) Quasi-particle spectrum shows that the superconducting $\frac{1}{4}$-hole doped strained graphene is gapless.}
\label{fig2}
\end{figure*}

\begin{figure*}
 \includegraphics[width=3.2in]{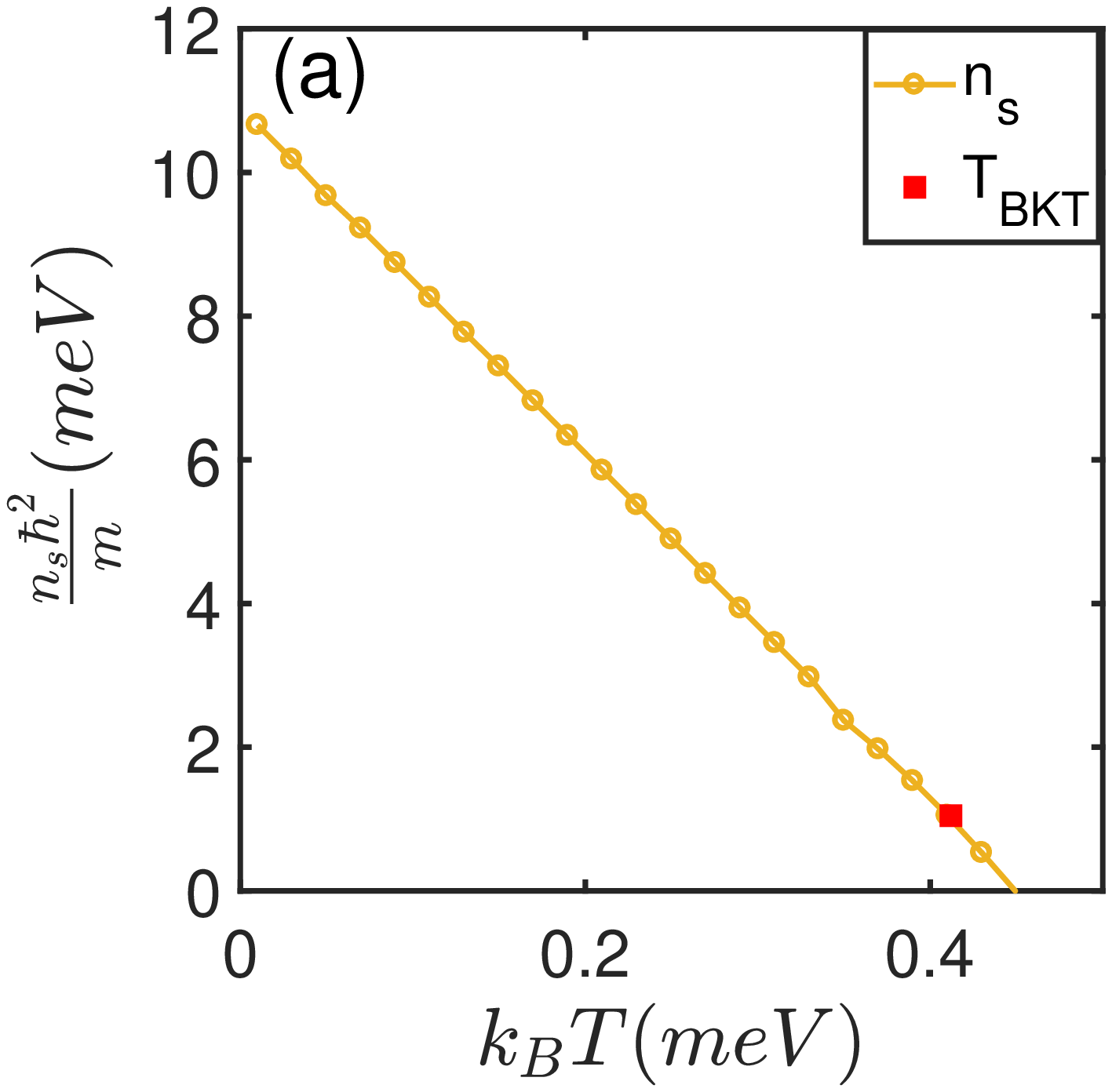}
 \includegraphics[width=3.2in]{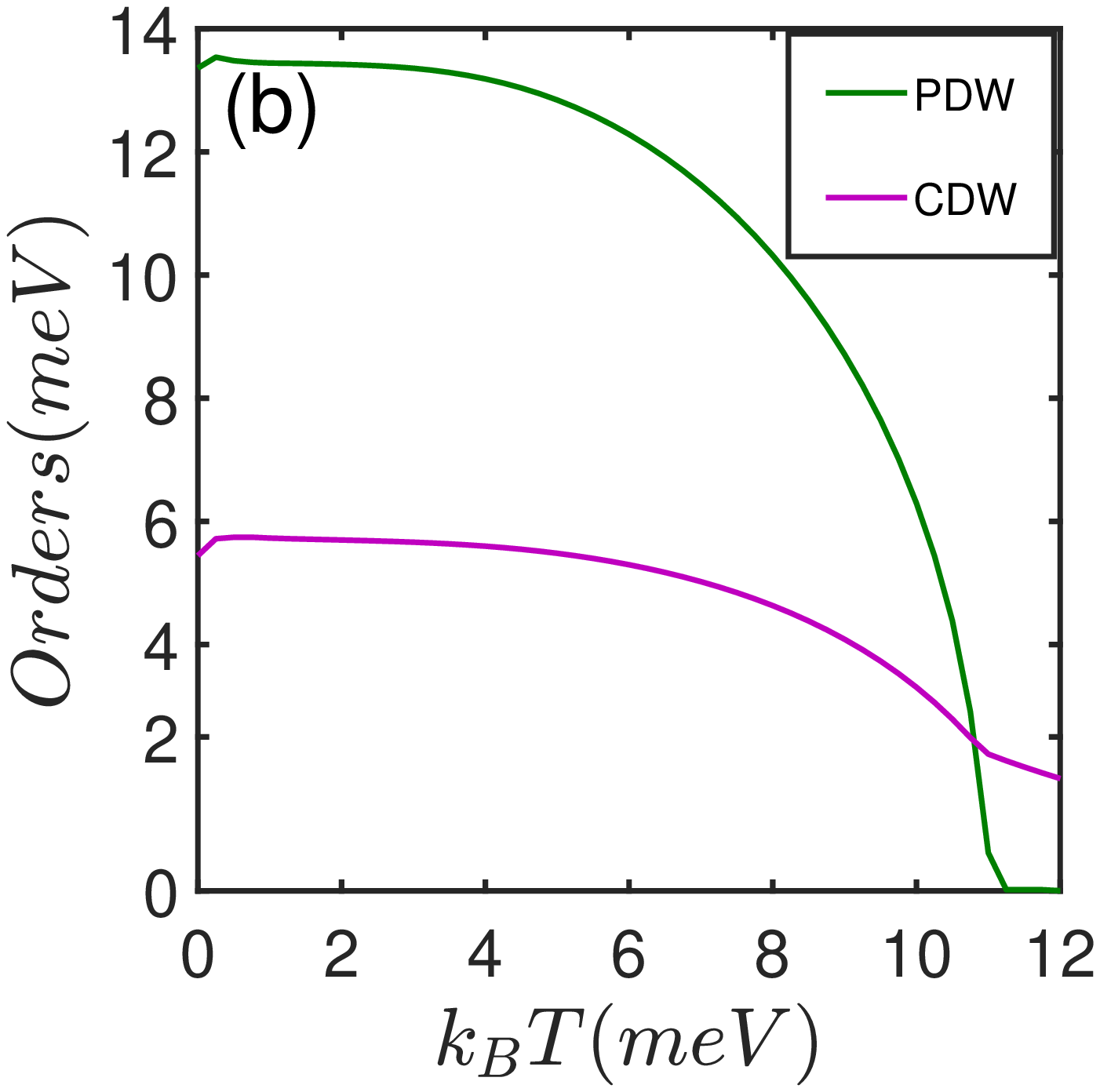}
\caption{(a)The superfluid weight of a pure superconducting pair-density-wave state in a lightly-doped strained graphene. Here parameters are $\delta =0.04$, $\mathbf{U}=4t$, $\frac{\delta t}{t}=0.05$, and $n=15$ . The line $\frac{8 k_BT}{\pi}$ is close to the temperature axis with the red dot representing the BTK transition temperature. (b) Following (a), the pair density wave order with momentum $\frac{Q}{2}$ and the charge density wave order as a function of temperature. }
\label{fig3}
\end{figure*}

\begin{figure*}
 \includegraphics[ width=3.2in]{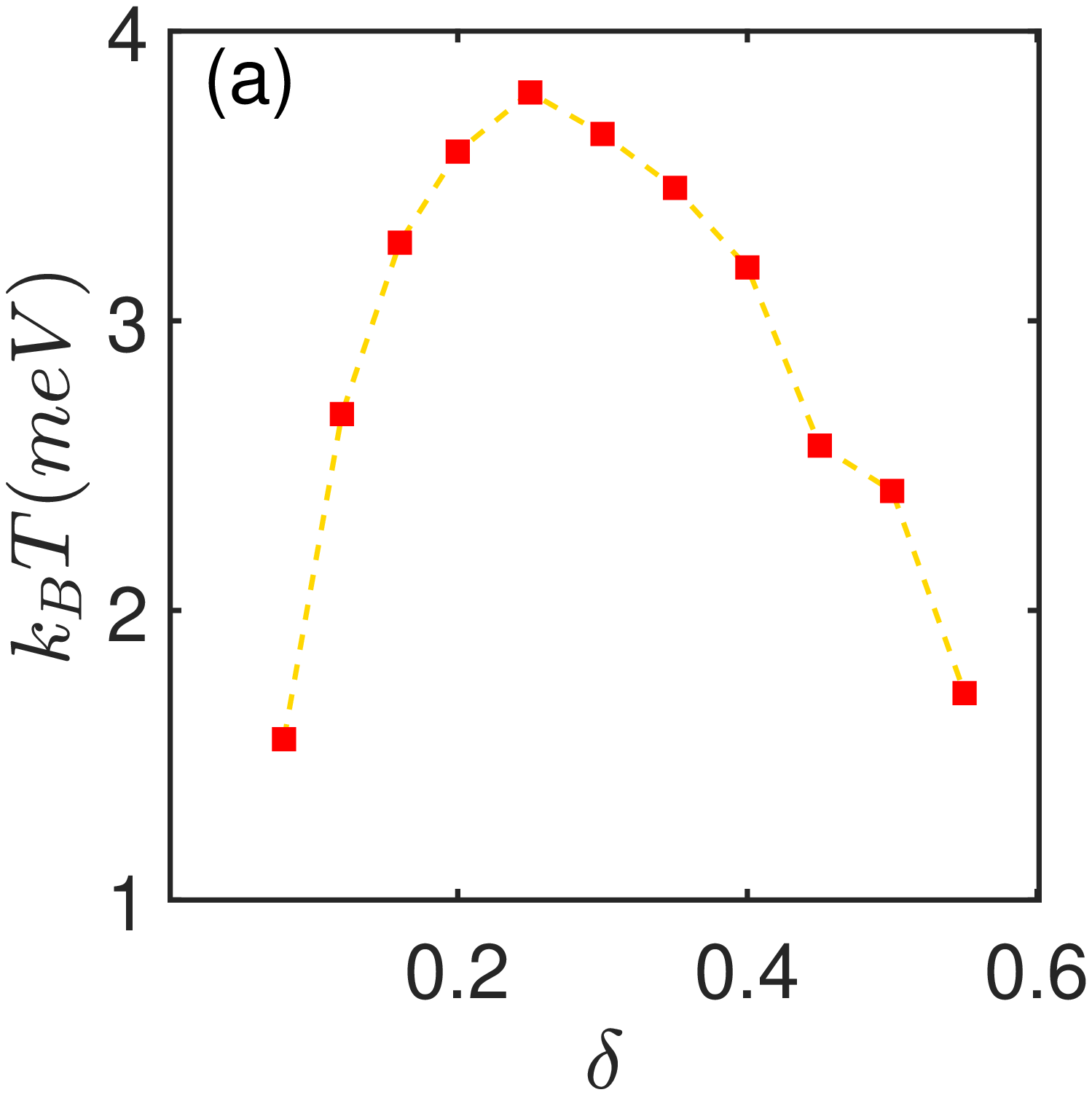}
  \includegraphics[width=3.2in]{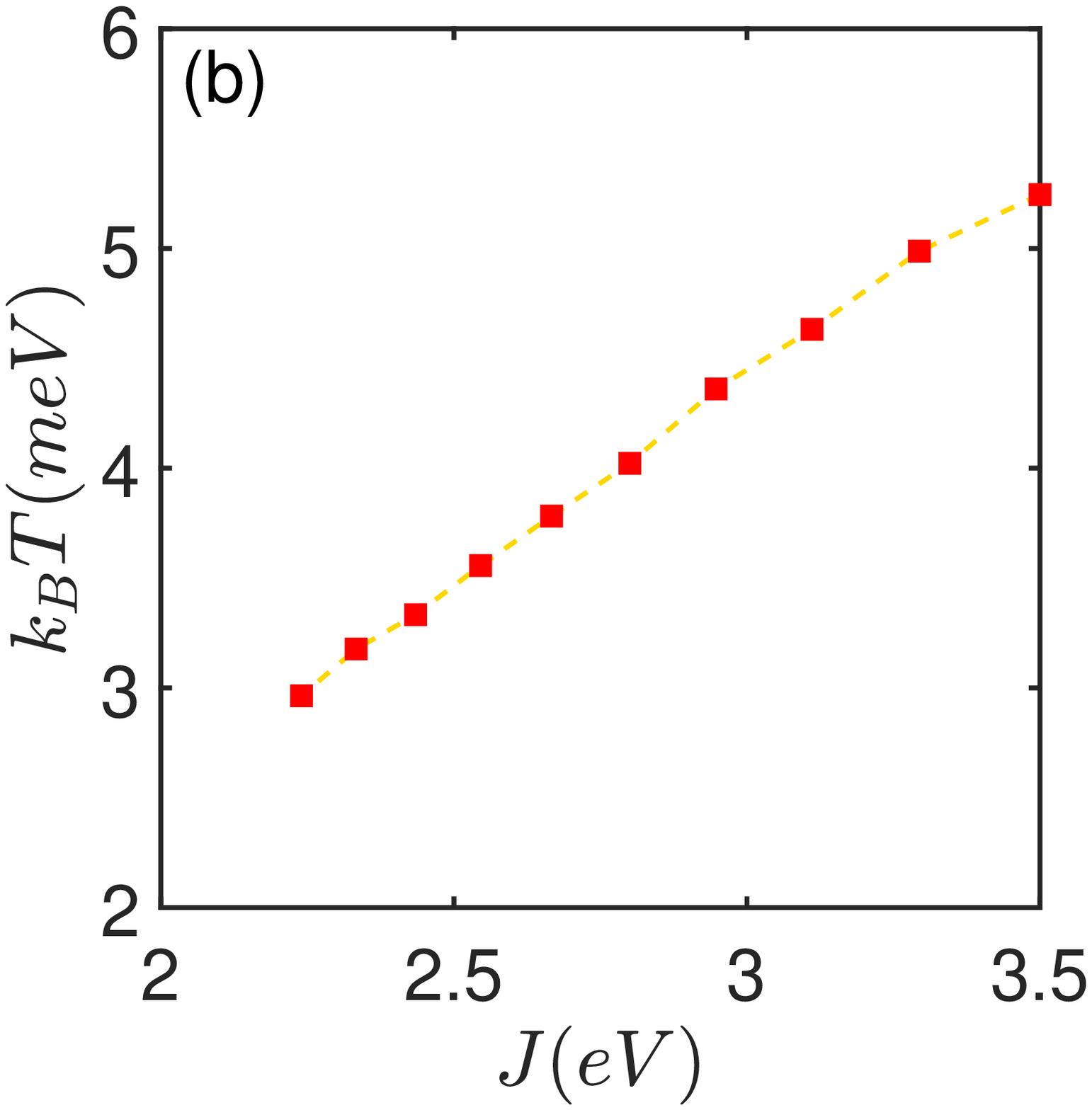}
\caption{(a)The BKT transition temperatures as a function of hole-doped level $\delta$ with strain amplitude being $\frac{\delta t}{t}=0.25$ and period being $n=15$. Here $\mathbf{U}=4t$ and the optimal hole doping density appears near $0.25$. (b) The BKT transition temperatures as a function of the strength of the spin-spin interaction with $\frac{\delta t}{t}=0.25$ and $n=15$. It shows an almost linear behavior, which is the same as the dependence of the transition temperature on the electron-phonon coupling when the BCS theory is specialized to a flat conduction band.}
\label{fig4}
\end{figure*}

\begin{figure*}
 \includegraphics[width=3.2in]{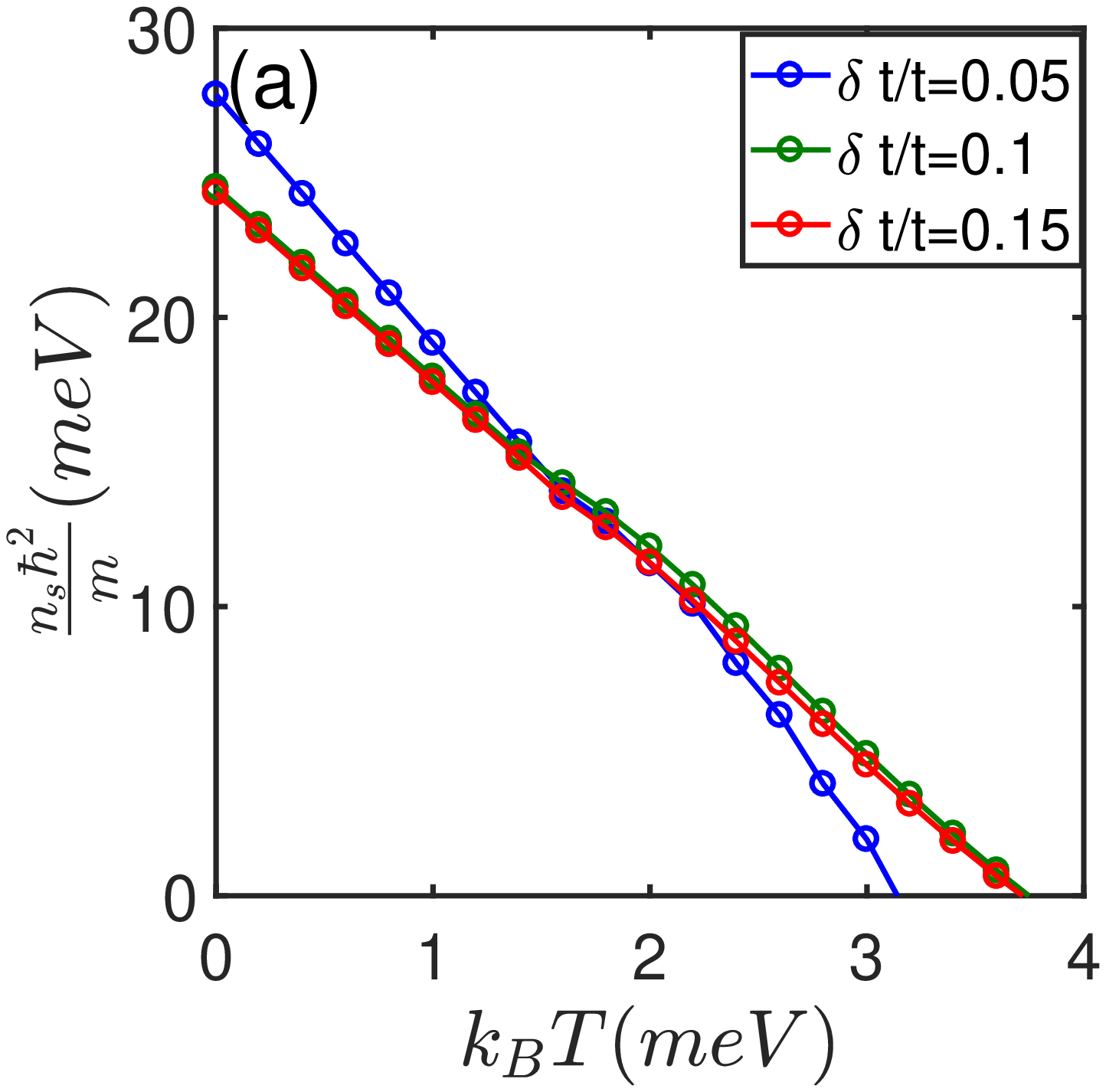}
 \includegraphics[width=3.2in]{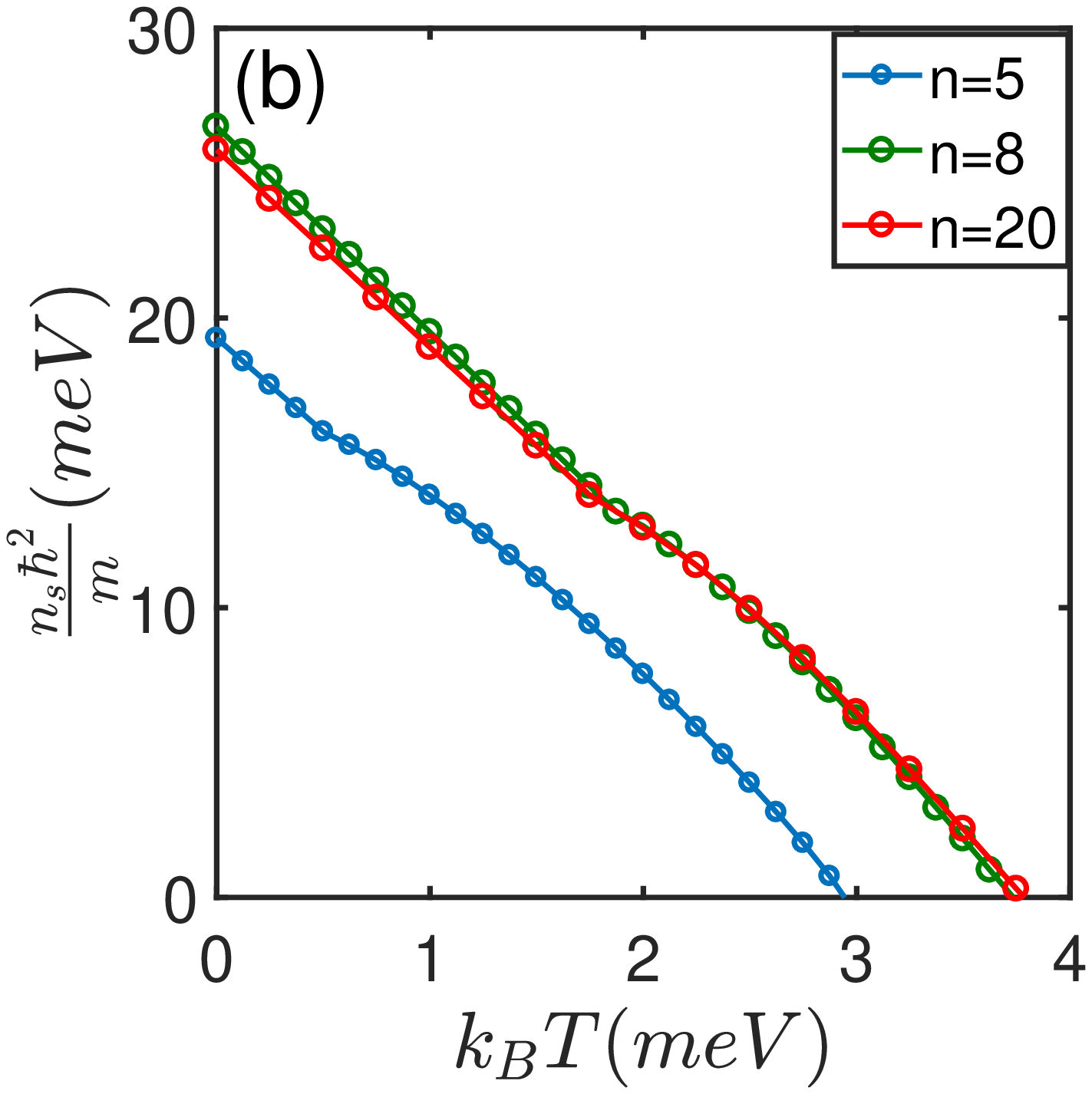}
\caption{TThe superfluid weight of the $\frac{1}{4}$-hole doped graphene with $\mathbf{U}=4t$ as a function of temperature for different amplitudes of the strain with $n=15$ (a) and periods of the strain with $\frac{\delta t}{t}=0.25$ (b).}
\label{fig5}
\end{figure*}
It is proposed that the chiral superconductivity could occur in doped graphene near the van Hove singularity at $\frac{1}{4}$ hole doping\cite{ref9}. Therefore, we shall first examine the possibility of superconductivity for  $\frac{1}{4}$ hole doped graphene in strong electron-electron interaction limit. In this case, the band energy dispersion exhibits a quadratic saddle point at the M-points in the resulting Fermi surface  when only the nearest-neighbor hopping is considered. Based on the Ginzburg-Landau theory, Nandkishore et. al argued that the chiral $d_{x^2-y^2}+id_{xy}$ combination of superconducting order that breaks the time-reversal and parity symmetry is the lowest energy in the pairing states\cite{ref5}. The ground state of the $\frac{1}{4}$-hole doped graphene with periodic strain is complicated due to the emergence of flat bands. It is reasonable that the superconducting energy gap $\Delta$ becomes position dependent and has the same period as that of the strain. This results in the emergence of the pair density wave along with the uniform chiral d-wave superconducting order, and simultaneously, the charge density wave emerging as a subsidiary order. Hence there are pair density wave order with momentum $Q$, the charge density wave order and the usual chiral d-wave superconductivity order in the ground state.  In this paper, we adopt a slightly different parameters with $n=15$ so that the phase of pair density wave occupies larger regime. Self-consistent solved results for strained graphene with $\frac{1}{4}$ doping to the van Hove singularity are given in Fig.~\ref{fig2}.  Here the charge density wave and the pair density wave orders coexist with the chiral d-wave superconductivity order in the strained graphene are compared with the pure chiral d-wave superconductivity order in the doped graphene without strain.  In Fig.~ \ref{fig2} (b),  we show the order parameters as a function of  temperature. It is seen that in the presence of strain, the surviving temperature for supercondcuting orders in strained graphene is much enhanced.  In the
Fig.~\ref{fig2} (a), we show the superfluid weight as a function of temperature. It is seen that the superfluid weight decreases with the temperature linearly. This is due to that in the pair density wave state, the center of mass for each Cooper pair carries momentum so that the energy of the quasi-particles is Doppler shifted and becomes gapless. This is shown in Fig.~\ref{fig2} (c).
The Berezinskii-Kosterlitz-Thouless transition temperature  is determined by the intersection of  $\frac{\hbar^2 n_s(T)}{m}$ with $\frac{8 k_BT}{\pi}$. As indicated in Fig.~\ref{fig2} (a), the BKT transition temperature is lower than the temperature when the superfluid weight disappears. The enhancement of the superconducting transition temperature with strain is clearly seen: The strain enhances the BKT  temperature from $25.35~K$  to  $37.70~K$ about increase $50\%$ in the $\frac{1}{4}$-hole doped graphene.

The pair density wave with momentum $Q$ is a trivial order induced by the periodic strain of period $L=2\pi/Q$.  It is therefore an exciting result that there is a pure superconducting pair-density-wave states (PDW) with momentum $\frac{Q}{2}$  accompanied by charge density wave along. Here there is no uniform chiral superconducting order.
The PDW state has been considered as a leading candidate to characterize the pseudogap state for the high-temperature superconductivity in cuprates. In Fig.~\ref{fig3}(a), we show how the superfluid weight of a pure PDW state changes as the temperature increases.  Here the doping level is $\delta=0.04$.
Clearly, it shows the same linear behavior as that for the normal chiral d-wave superconducting state with $\Delta_Q$ being in presence. The line $\frac{8 k_BT}{\pi}$ is close to the temperature axis with the red dot representing the BTK transition temperature, which is about $5~K$. The thermal evolution of orders versus temperature is shown in the Fig.~\ref{fig3}(b). The gap opening temperature is seen to be $T^* \sim 120~K$, which is much higher than the superconducting critical temperature $T_{BTK} \sim 5~K$ in the lightly hole doped strained graphene. The difference between $T^*$ and $T_{BTK}$ is similar to the difference in the pseudogap temperature and the superconducting transition temperature in the high-temperature cuprate superconductors and may be used to understand the origin of the pseudogap state in cuprate superconductors.

In Fig.~\ref{fig4}(a), we show the BTK transition temperature versus the doping level $\delta$. Clearly, in resemblance to the superconducting dome observed in high-$T_c$ cuprate superconductors, it also exhibits a dome-like shape. We can therefore divide the superconducting region into underdoped region, the optimal doping, and overdoped region. The optimal doping level is at $\delta=\frac{1}{4}$ where the van Hove singularity occurs in the band structure of unstrained graphene. The BKT transition temperatures increase with the doping level in the underdoping region, while it decrease with the doping level in the overdoped region. The highest superconducting transition temperature is about $37.70~K$ in the case of $\delta=\frac{1}{4}$ with the Hubbard interaction $\mathbf{U}=4t$. The BKT transition temperatures for different strong Hubbard interactions from $\mathbf{U}=3t$ to $\mathbf{U}=5t$ are shown in the Fig.~\ref{fig4}(b). It is obvious that the BKT  transition temperatures increases linearly with the residual spin-spin interaction strength $J=\frac{4t^2}{\mathbf{U}}$ even in the strong correlation limit. This is the same as the dependence of the transition temperature on the electron-phonon coupling when the BCS theory is specialized to a flat conduction band. It is a characteristics of the flat band superconductivity in contrast to the relation for the critical temperature $T_c \sim e^{-\frac{1}{g}}$ in the conventional superconductors. The flat band superconductivity is thus a potential route to high-temperature superconductivity even in the presence of strong and repulsive correlation effect.

Finally, we discuss effects of the period ($n$) and amplitude ($\delta t$) of the strain on the superfluid weight for the $\frac{1}{4}$ hole-doped graphene with strain. As shown in the Fig.~\ref{fig5}(a), we see that the superfluid weights for large amplitudes of strain (greater than $0.1$) are almost the same near the zero temperature.  As a result, the BKT transition temperature slightly increases with the creasing of the amplitude in strain beyond $0.1$. Due to the periodicity imposed by the period $n$, the flat bands in the k space would repeat itself with the period $\frac{4\pi}{3na}$ in $k_x$ direction. This is similar to the higher harmonics of the pair density wave $\Delta_{nQ}$ with $n=1,2,3, \cdots$. Furthermore, flat bands get flatter as the period of strain increases.  As a result, for $n>8$, the superfluid weights are almost the same with $n=8$ as shown in the Fig.~\ref{fig5}(b). However, it should be noted that as $n$ increases, there are more flat bands, which result in higher BKT transition temperatures.

\section{Conclusions}
\label{sec:con}

In summary, we have shown that due to the presence of flat bands, the superconducting transition temperature for the chiral $d$ wave superconductivity in the doped graphene with strain is enhanced  approximately $50\%$ in comparison to that of the doped graphene without strain. We also show that for the unconventional superconducting state that coexists with the charge density wave and the pair density wave in strained graphene,  the superfluid weight as a function of temperature exhibits the same trend as that for pure the chiral $d$ wave superconductivity in the unstrained graphene. The non-vanishing superfluid weight of pure pair density wave state in the lightly doped graphene demonstrates that it is a superconducting state with the critical temperature (the BTK temperature) being much lower than the gap-opening temperature. This phenomenon can be used to understand the origin of the  pseudogap state observed in the high-temperature cuprate superconductors. In addition, we find that as the period and amplitude of the strain increases,  the superfluid weight will saturate once flat bands emerge. Finally, we show that in resemblance to the superconducting dome observed in high-$T_c$ cuprate superconductors\cite{ref13} and twisted bilayer graphene\cite{ref12}, the BKT transition temperature versus doping for strained graphene also exhibits a dome-like shape. However, unlike the exponential dependence of $T_c$ on the spin-spin interaction strength in high-$T_c$ cuprate superconductors, the BKT transition temperature of strained graphene depends linearly on the spin-spin interaction strength.

 \section{Acknowledgment}
This work was supported by the National Science Foundation of China (Grant Nos:11804213), Scientific Research Program Funded by Shaanxi
Provincial Education Department (Grant Nos:20JK0573) and Scientific Research Foundation of Shaanxi University of Technology (Grant Nos: SLGRCQD2006). C. Y. Mou was supported by the Ministry of
Science and Technology (MoST), Taiwan. He was also supported by the Center for Quantum Technology within the framework
of the Higher Education Sprout Project by the Ministry of Education (MOE) in Taiwan.

\end{CJK*}
\end{document}